\documentclass[aps,prl,twocolumn,superscriptaddress,letterpaper]{revtex4-2}

\usepackage{graphicx}
\usepackage{bm}
\usepackage{amssymb}
\usepackage{color}
\usepackage{amsmath}
\usepackage{ulem}

\begin{document}

\title{Topological slow light via coupling chiral edge modes with flat bands}

\author{Letian Yu}
\affiliation{Division of Physics and Applied Physics, School of Physical and Mathematical Sciences, Nanyang Technological University,
Singapore 637371, Singapore}

\author{Haoran Xue}
\email{haoran001@e.ntu.edu.sg}
\affiliation{Division of Physics and Applied Physics, School of Physical and Mathematical Sciences, Nanyang Technological University,
Singapore 637371, Singapore}

\author{Baile Zhang}
\email{blzhang@ntu.edu.sg}
\affiliation{Division of Physics and Applied Physics, School of Physical and Mathematical Sciences, Nanyang Technological University,
Singapore 637371, Singapore}
\affiliation{Centre for Disruptive Photonic Technologies, Nanyang Technological University, Singapore, 637371, Singapore}

\begin{abstract}
Chiral edge modes in photonic topological insulators host great potential to realize slow-light waveguides with topological protection. Increasing the winding of the chiral edge mode around the Brillouin zone can lead to broadband topological slow light with ultra-low group velocity. However, this effect usually requires careful modifications on a relatively large area around the lattice edge. Here we present a simple and general scheme to achieve broadband topological slow light through coupling the chiral edge modes with flat bands. In this approach, modifications inside the topological lattice are not required. Instead, only several additional resonators that support the flat bands need to be attached at the lattice edge. We demonstrate our idea numerically using a gyromagnetic photonic crystal, which is ready to be tested at microwave frequencies.
\end{abstract}

\maketitle

Topological photonics \cite{lu2014,khanikaev2017,ozawa2019}, a rapidly growing research field that studies the physics of topological phases \cite{hasan2010,qi2011} in electromagnetic systems, provides powerful platforms for new devices with unconventional functionalities and superior performance. Over the last decade, various topological phases have been proposed and realized in photonics, such as Chern insulators \cite{haldane2008,raghu2008,wang2008,wang2009}, Floquet topological insulators \cite{fang2012,rechtsman2013,liang2013,gao2016,maczewsky2017,mukherjee2017}, qunatum-spin-Hall-like topological insulators \cite{hafezi2011,hafezi2013,khanikaev2013,chen2014,wu2015,cheng2016}, valley-Hall topological phases \cite{ma2016,dong2017,wu2017,gao2018,noh2018} and topological semimetals \cite{lu2013,lu2015}. In particular, chiral edge modes (CEMs) in photonic Chern insulators, which are protected without any symmetries and are robust against any kinds of gap-preserving perturbations, are promising for potential applications. One natural usage of the CEMs is to build topological slow-light waveguides. Slow light has wide applications ranging from enhancing light-matter interaction to optical buffering \cite{baba2008}, but is sensitive to fabrication imperfections which causes undesired backscattering and localization \cite{melati2014}. Thus, encoding topological protection in slow-light devices would greatly enhance the performance.

Bulk-boundary correspondence states that the existence of the CEMs is solely determined by the Chern number of the bulk bands \cite{thouless1982}. Modifications near the edge will not affect the existence of the gapless CEM but only change its dispersion curve. Thus, a proper modification may produce topological slow light. In fact, one can easily obtain a CEM with very small slope over a narrow frequency window (compared to the bandgap size) by tuning the geometry of the edge \cite{yang2013,lan2020}. However, achieving broadband topological slow light, which is crucial for many applications \cite{baba2008}, is a challenging task. To overcome this narrow bandwidth issue, Guglielmon and Rechtsman recently proposed the idea of letting the CEM wind multiple times around the Brillouin zone (BZ), leading to broadband topological slow light \cite{guglielmon2019}. However, this usual edge dispersion requires complicated modifications around the edge. In the lattice model presented in Ref.~\cite{guglielmon2019}, both nearest-neighbor couplings and next-nearest-neighbor couplings need to be tuned according to their positions. Moreover, a higher-winding of the CEM requires a deeper modification into the bulk. It would be highly desired if a simpler method to achieving higher-winding CEM can be proposed.

In this Letter, we present a simple and general scheme to achieve CEM with higher momentum-space winding for broadband topological slow light. Our idea is based on coupling the CEM with flat bands, which are supported by additional sites attached to the lattice edge. We demonstrate our idea using gyromagnetic photonic crystals at microwave frequencies. We show that the number of times that the CEM winds around the BZ can be controlled by the number of flat bands available inside the bandgap. Under higher momentum-space winding, topological slow light with large bandwidth was observed. Moreover, an enlarged bulk area is not required to realize this phenomenon since the modifications are only around the lattice edge.

The proposed scheme is schematically illustrated in Fig.~1(a), where the grey rectangle represents a photonic Chern insulator. Here the detailed configurations, like the lattice geometry, are not important. At the edges of the photonic Chern insulator, there is a CEM that propagates unidirectionally (say the gap Chern number $C_{\text{gap}}=1$). Without additional edge modification, a natural edge termination \cite{wang2008,wang2009,poo2011,yang2013} is most likely to support a CEM that does not wind around the BZ (see the red curve in Fig.~1(b)). In addition to the Chern insulator, there is an array of resonators placed near the bottom edge of the Chern insulator, as indicated by the blue circles in Fig.~1(a). Each of these resonators support some resonances lying inside the topological bandgap of the Chern insulator. Later we will show how this can be achieved in one realistic setting. When these resonators are isolated from each other and also from the Chern insulator, the corresponding dispersions are nothing but just some flat bands as denoted by the blue lines in Fig.~1(b). Then we introduce couplings between these flat bands and the CEM by physically connecting the resonators to the edge of the Chern insulator. Now the new edge dispersion, which results from the hybridization between the original CEM and the flat bands, winds multiple times across the one-dimensional BZ (see Fig.~1(c)). Since the group velocity is the slope of the dispersion ($v_g=dw/dk$), such multiple winding of the edge dispersion naturally leads to very low group velocity. There are several advantages which can be directly seen from above description. First, this approach does not require any changes in the topological lattice itself. Thus, only a relatively small lattice with a well-defined bulk is required. Second, this approach is general since no detailed information on the topological lattice is required. In principle, it works for different systems with different lattices or different working frequencies, as long as a CEM is presented. Third, the construction of such a structure would be relatively simple. To have multiple flat bands inside the bandgap, we can, for example, use a ring resonator with several resonance modes inside the bandgap or stack several resonators along $z$ direction and each of them support one resonance mode inside the bandgap.

\begin{figure}
  \centering
  \includegraphics[width=\columnwidth]{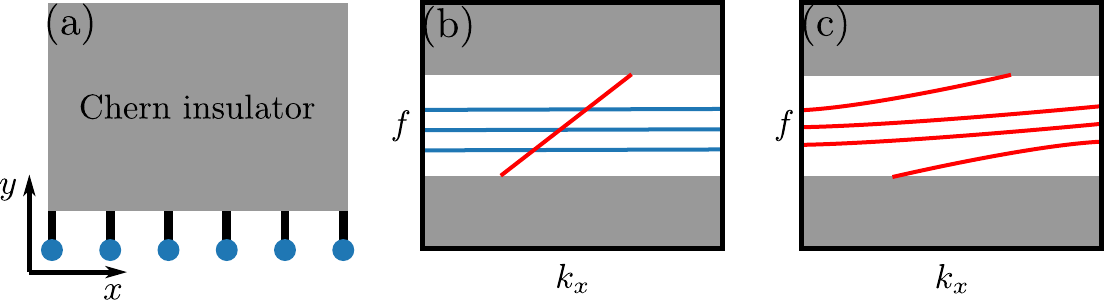}
  \caption{Constructing topological slow light via coupling CEM with flat bands. (a) shows the schematic of the structure. The grey rectangle represents a photonic Chern insulator with CEM propagating unidirectionally along the edges. The blue circles at the bottom denote resonators which support resonances with frequencies insides the bulk bandgap of the Chern insulator. When these resonators are isolated from the environment, they give rise to several flat bands (blue lines in (b)) inside the bandgap which do not interact with the CEM (red curve in (b)). When coupling channels between the resonators and lattice edge are enabled, the flat bands hybridize with the CEM, leading to a new CEM with higher momentum-space winding as shown in (c).}
  \label{fig1}
\end{figure}

\begin{figure}[t]
  \centering
  \includegraphics[width=\columnwidth]{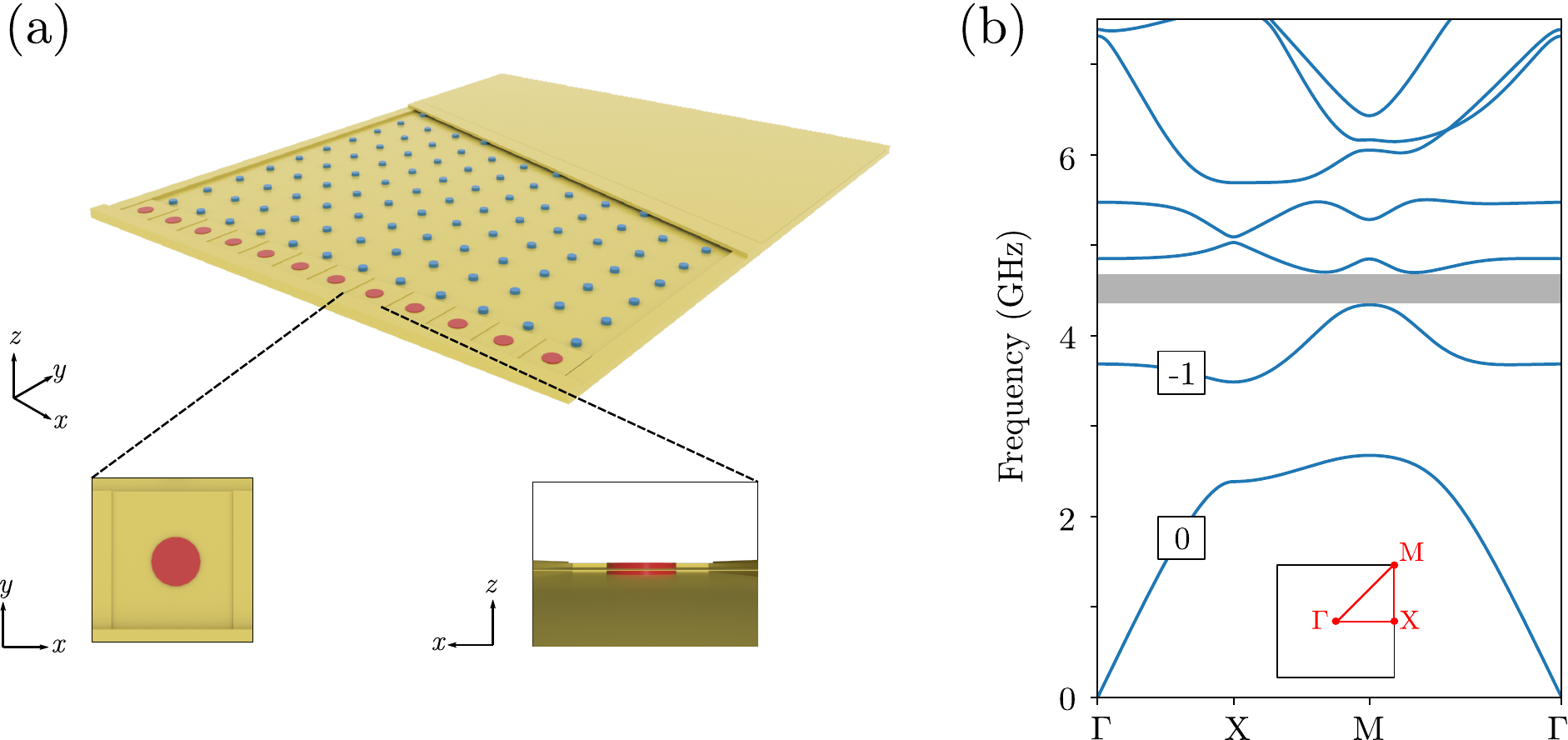}
  \caption{Realistic photonic structure for slow light waveguide and its corresponding photonic crystal band structure. (a) shows schematic of the waveguide composed of gyromagnetic photonic crystal rods (blue rods) and metal boundaries (yellow). The rods are confined within metal plates and the empty space inside is filled with air. Here, we stacked $2$ resonators along $z$ direction for each unit cell on the lower edge, which corresponds to the case of winding $2$ times around the BZ. The lower panel of (a) consists of enlarged top and side views of boundary modification. Note that there should be a top metal plate covering these additional resonators, which is omitted here. In (b), we present transverse magnetic band structure of the square lattice obtained using full-wave simulations. Topological band gap supporting CEM, which is used for boundary modifications, is shaded in grey.}
  \label{fig2}
\end{figure}

Now we proceed to design a realistic structure using the scheme described above. To begin with, we need a photonic topological lattice with CEM. Here we used gyromagnetic photonic crystals, which are widely adopted to realize photonic Chern insulators at microwave frequencies \cite{wang2008,wang2009,poo2011,yang2013}, to demonstrate our idea. Consider a square lattice consisting of cylindrical rods sandwiched by two metal plates, as shown in Fig.~2(a). These rods are made of gyromagnetic material with $\epsilon=14.63\epsilon_0$. Under a static magnetic field that is perpendicular to the lattice, time-reversal symmetry is broken and the permeability tensor takes the following form:
\begin{equation}
	\bm{\mu}=\left(
	\begin{array}{ccc}
		\mu & i\kappa & 0\\
		-i\kappa & \mu & 0\\
		0 & 0 & \mu_0\\
	\end{array}\right).
\end{equation}
Throughout this paper, we have set $\mu=14\mu_0$ and $\kappa=12.4\mu_0$. Other parameters for the bulk lattice are: lattice constant $a=40$ mm, radii of the rods $r=3.9$ mm and heights of the rods $h=4.5$ mm. The bulk band structure for transverse magnetic polarization obtained from full-wave simulation is plotted in Fig.~2(b). The Chern number for each band is defined as: $C=\frac{1}{2\pi}\iint_{\text{BZ}}(\frac{\partial A_y}{\partial x}-\frac{\partial A_x}{\partial y})d^2\bm{k}$, where $\bm{A}=-i\langle \bm{u}_n|\partial_{k}|\bm{u}_n\rangle$ is the Berry connection with $\bm{u}_n$ being the periodic part of the Bloch wavefunction of $n$th band. The Chern numbers for the first two bands are 0 and -1, respectively. Thus, the second bandgap (from 4.34 GHz to 4.70 GHz, shaded in grey in Fig.~2(b)) is topologically nontrivial, supporting one unidirectional CEM. 

Having designed the photonic Chern insulator and identified its topological bandgap that support one CEM, we now turn to construct the flat bands to couple with the CEM. To this end, we put an additional array of rods at the edge of the lattice (see the insets of Fig.~2(a) for details). These rods are made of dielectric with $\epsilon=4.0\epsilon_0$, and each of them support a resonance mode within the topological bandgap. As can be seen from the lower panel of Fig.~2(a), these dielectric rods are stored in individual hollow boxes made by metals (acting as perfect electric conductor (PEC) at microwave regime) and only couple to the edge of the lattice. Thus, this configuration realizes the idea of coupling CEM with flat bands. Moreover, by controlling the number of dielectric rods that stack along $z$ direction, we can control the number of flat bands which couple with the CEM. In the case shown in Fig.~2(a), there are two dielectric rods along $z$. This number can be easily increased by stacking more rods along $z$. Note the radii of these rods at different layers should be slightly different to produce resonances at different frequencies inside the bandgap.

\begin{figure}
  \centering
  \includegraphics[width=\columnwidth]{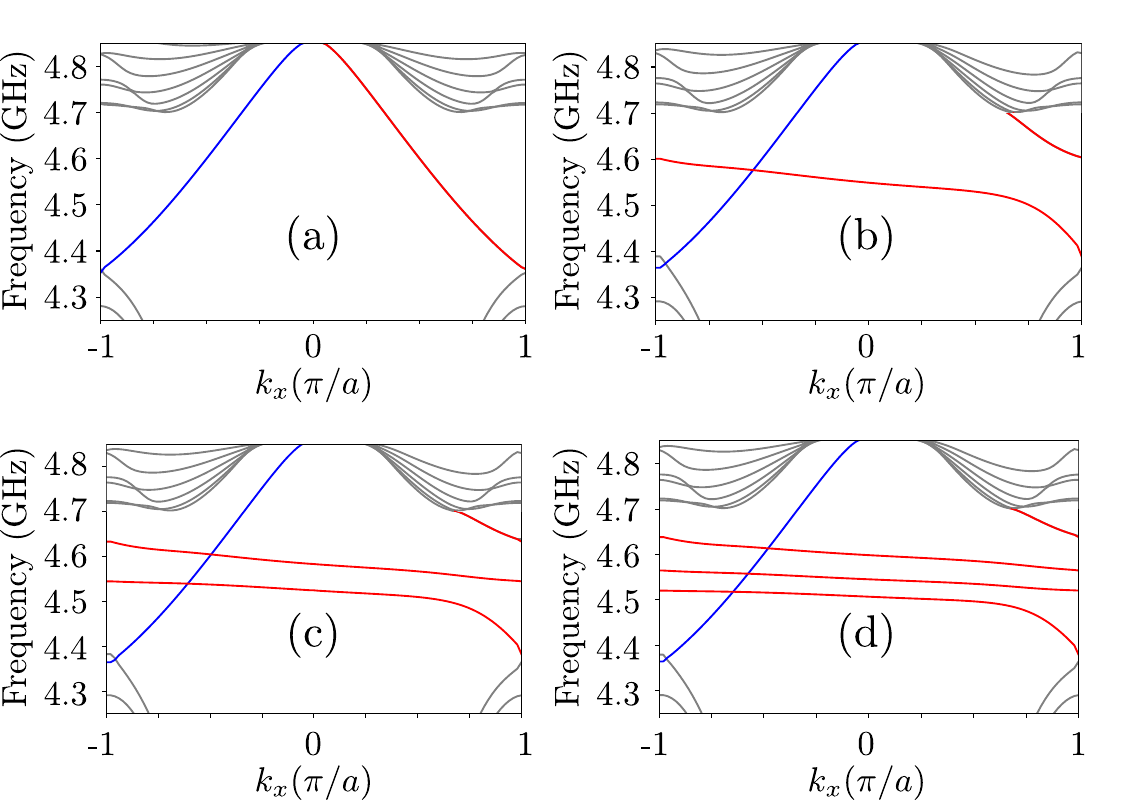}
  \caption{Band structures for different number of additional resonators along $z$ direction. Panels (a)-(d) show the results obtained by using $n=0,1,2,3$ resonators to couple with the bottom-localized CEM. The top-localized edge mode is coloured blue and the bottom-localized edge mode, which is coupled to the additional resonators, is coloured red. For clarification purpose, the top-localized edge mode was left un-treated and all boundary modifications were performed on the lower lattice edge.}
  \label{fig3}
\end{figure}

To see how actually our design works, we performed full-wave simulations to obtain the edge dispersions under different boundary modifications. In the simulations, we used structures that were finite along $y$ direction and periodic along $x$ direction. We have taken 10 unit cells along $y$ direction to ensure there are no couplings between the edge states on opposite edges. The top edge was simply terminated by a PEC while the bottom edge was decorated by additional resonators as illustrated in Fig.~2(a). The number of additional resonators are denoted by $n$ (The detailed structural parameters for different $n$ are given in Supplementary Materials \cite{SM}). When $n=0$ (i.e., no additional resonators are supplied and the lattice is simply terminated by a PEC), we saw two gapless edge modes inside the bulk bandgap (Fig.~3(a)). The one colored in red corresponds to the CEM on the bottom edge. It is obvious from Fig.~3(a) that this edge mode does not wind around the first BZ. Next, we put one additional resonator on the edge to couple with the CEM. The resulted dispersion, as given in Fig.~3(b), shows that now the edge mode (red curve) winds around the BZ once due to the presence of the additional resonator. Similarly, for $n=2$ and $n=3$ (i.e., two and three additional resonators), the edge modes wind around the BZ for two (Fig.~3(c)) and three (Fig.~3(d)) times, respectively. These results demonstrate that coupling CEM with flat bands can indeed induce higher winding of the edge mode and the winding number is controlled by the number of flat bands (or additional resonators) that are coupled with the CEM.

\begin{figure}
  \centering
  \includegraphics[width=\columnwidth]{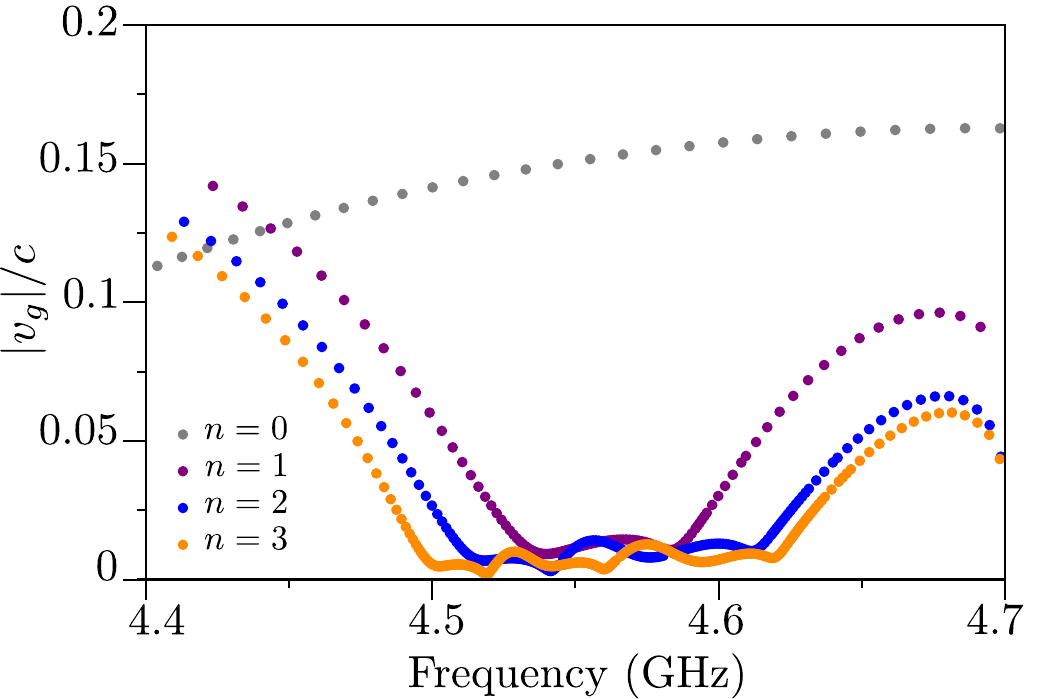}
  \caption{Group velocity for the bottom-localized edge mode plotted as a function of frequency. Individual curve corresponds to each of the band structures in Fig.~3. Different $n$ values are represented by different colours as indicated in the figure. All data points here were obtained from full-wave simulations performed in Fig.~3. Except for the trivial case $n=0$, all other curves have a clear 'frequency window', in which the group velocity keeps low and remains approximately constant.}
  \label{fig4}
\end{figure}

So far we have achieved higher winding of the edge mode by putting additional resonators on the edge. However, such higher winding property alone is not sufficient to guarantee the existence of practically useful topological slow light. It only indicates a general reduction of group velocity inside the bandgap. Thus, we have to check the exact relation between frequency and the group velocity of the edge mode. The group velocity can be directly extracted from the edge dispersion given in Fig.~3 through $v_g=dw/dk_x$. The calculated group velocities of the CEM for different $n$ ($n=0,1,2,3$) are plotted in Fig.~4. There are several important features that be read from Fig.~4. Frist, a higher winding ($n=1,2,3$) leads to a reduction of group velocity for most frequencies inside the bandgap, except for the frequencies near the lower band edge. Second, the group velocity is negative for all frequencies inside the bulk bandgap (i.e., the edge dispersion is monotonically decreasing, ensuring $v_g<0$). Third, for $n=1,2,3$, there exists a frequency window in which the group velocity is quite small and the variation of the group velocity is also small. This frequency window is essential for broadband topological slow light and its width increases with $n$. We call this frequency window the broadband topological slow light region. For $n=3$, it occupies around 34\% of the bandgap. This region can be further widened by increasing $n$ (i.e., adding more resonators on the edge). Considering all these features described above, our structure provides a concrete example for generating broadband topological slow light at microwave frequencies based on the scheme of coupling CEM with flat bands. We note that it does not require a complicated optimization process to obtain the results presented in Fig.~3 and Fig.~4. Instead, one only needs to tune the radii of the additional resonators to adjust their frequencies and the size of the coupling channel to the lattice edge to tune the coupling strength with the edge mode (see Supplementary Materials \cite {SM} for more details).  

In conclusion, we have presented a simple and effective approach for generating broadband topological slow light through coupling CEM with flat bands. A realistic design based on a gyromagnetic photonic crystal is presented to show the feasibility of our general approach. While our example is at microwave frequencies, in principle the idea can also be implemented in optical frequencies. Although for on-chip optics stacking resonators along out-of-plane direction may not be feasible, additional flat bands can be supplied by other means like using a single resonator with multipole resonances inside the topological gap. To further enhance the performance of the topological slow light, it may also be a good idea to combine the scheme introduced here together with a optimization process which can yield a larger bandwidth and less group velocity variation. Besides, this approach can also be used in other classical-wave systems such as acoustics \cite{yang2015,ni2015,khanikaev2015,ding2019} where Chern insulators can be implemented.\\

This work was supported by the Singapore MOE Academic Research Fund Tier 3 Grant MOE2016-T3-1-006 and Tier 2 Grant MOE 2018-T2-1-022(S).

%

\clearpage
\newpage
\onecolumngrid
\begin{center}
  \textbf{\large Supplementary Materials}\\[.2cm]
\end{center}
\twocolumngrid
This section discusses detailed parameters used for results obtained in Fig.~3 and Fig.~4. In addition to bulk's parameters (covered in main section), additional resonators' frequencies and coupling strengths to the chiral edge mode would change the shape of the hybridized edge mode. We would illustrate how to tune these two factors.

For the additional resonator, we have chosen it to be a cylinder residing in an empty box filled with air. The cylinder is made of materials with $\epsilon=4\epsilon_0$ and its height is $0.5$ mm. The box has a dimension of $36\times36\times0.5$ mm$^3$. To prevent neighboring resonators from coupling with each other, metal blocks were attached to faces of the box which act as perfect electric conductor boundary conditions. Only the face facing the lattice was remained open to allow couplings to take place. 

\begin{figure}[h]
  \centering
  \includegraphics[width=\columnwidth]{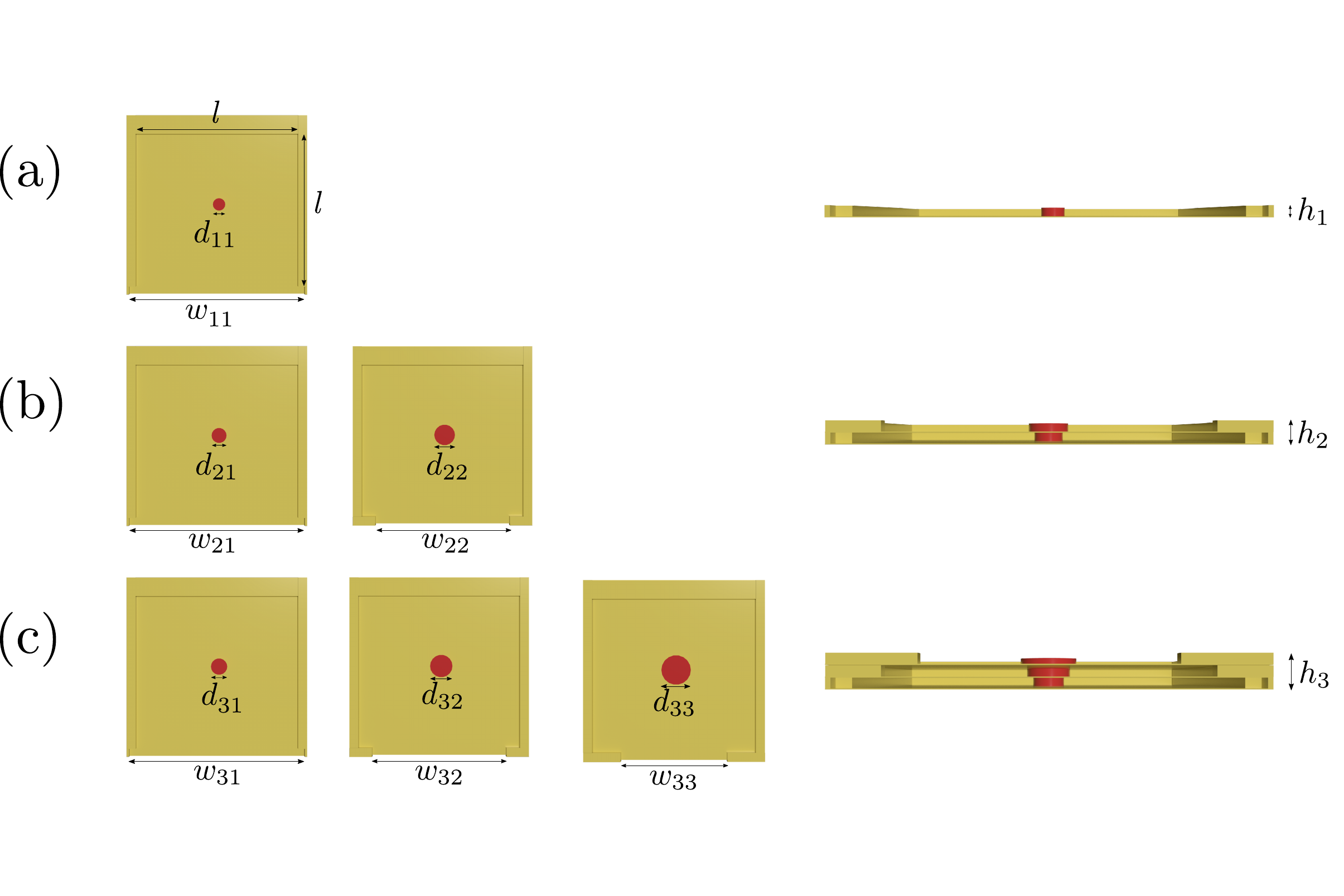}
  \caption{Cross-sectional views and side views for different boundary modifications. (a) to (c) correspond to the cases of $n=1,2,3$ respectively. The cylindrical gyromagnetic materials are in red and the metal blocks are in yellow. Here subscripts $ab$ on diameter $d$ and channeling width $w$ refer to the number of additional resonators and the position along $z$ direction, respectively. The left panel draws cross-sectional views of the additional resonators in scale. For individual case from left to right, the resonators were drawn in the order of bottom to top. With everything in millimeter, the parameters are $d_{11}=2.74$, $w_{11}=39$, $d_{21}=3.3$, $w_{21}=39$, $d_{22}=4.6$, $w_{22}=30$, $d_{31}=3.6$, $d_{32}=5.0$, $d_{33}=6.5$, $w_{31}=39$, $w_{32}=30$ and $d_{33}=23.5$. The right panel depicts side views of the additional resonators, with $h_1=0.5$, $h_2=1.1$ and $h_3=1.7$.}
  \label{fig5}
\end{figure}

There are mainly two factors affecting the coupling strengths between the additional resonators and the lower edge of the lattice. One being the size of the resonators' face facing the lower edge and the other being the distance between these two sites. For the face facing the lower edge, if part of it is covered by metal blocks, this would result in lower coupling strength between additional resonator and the lattice sites. In subsequent paragraphs, we will denote this covering by the width of the remaining unblocked area (see Fig.~5) and call this as 'channeling width' for the additional resonators. Throughout simulations, we have decided to fix the distance from the additional resonators to the bottom of the lattice to be 1 mm. In adjusting distance between the addtional resonators and the lower edge, we have trimmed $5$ mm from the periodic lattice's lower edge (so each bottom unit cell has a size $35\times40$ mm$^2$ instead of  $40\times40$ mm$^2$) for all simulations except for the case $n=0$.  

All simulations were performed with COMSOL Multiphysics. Every additional resonator consists of a cylindrical rod and an empty square air box of side length $l=36$ mm. For $n=1$, the cylindrical rod has diameter $d_{11}=2.74$ mm and channelling width $w_{11}=40$ mm. For $n=2$, the bottom rod has diameter $d_{21}=3.3$ mm with channelling width $w_{21}=40$ mm. The top rod has diameter $d_{22}=4.6$ mm with channeling width $w_{22}=30$ mm. For $n=3$, from bottom to top, we have $d_{31}=3.6$ mm, $d_{32}=5.0$ mm, $d_{33}=6.5$ mm and  $w_{31}=40$ mm, $w_{32}=30$ mm, $d_{33}=23.5$ mm. In our simulations, we separated additional resonators along $z$ direction using $s=0.1$ mm metal plate, which could be changed to any non-zero value as its sole purpose was to prevent couplings in between additional resonators. In the right panel of Fig.~5, we have $h_1=0.5$ mm, $h_2=1.1$ mm and $h_3=1.7$ mm. 


\begin{thebibliography}{40}%
\makeatletter
\providecommand \@ifxundefined [1]{%
 \@ifx{#1\undefined}
}%
\providecommand \@ifnum [1]{%
 \ifnum #1\expandafter \@firstoftwo
 \else \expandafter \@secondoftwo
 \fi
}%
\providecommand \@ifx [1]{%
 \ifx #1\expandafter \@firstoftwo
 \else \expandafter \@secondoftwo
 \fi
}%
\providecommand \natexlab [1]{#1}%
\providecommand \enquote  [1]{``#1''}%
\providecommand \bibnamefont  [1]{#1}%
\providecommand \bibfnamefont [1]{#1}%
\providecommand \citenamefont [1]{#1}%
\providecommand \href@noop [0]{\@secondoftwo}%
\providecommand \href [0]{\begingroup \@sanitize@url \@href}%
\providecommand \@href[1]{\@@startlink{#1}\@@href}%
\providecommand \@@href[1]{\endgroup#1\@@endlink}%
\providecommand \@sanitize@url [0]{\catcode `\\12\catcode `\$12\catcode
  `\&12\catcode `\#12\catcode `\^12\catcode `\_12\catcode `\%12\relax}%
\providecommand \@@startlink[1]{}%
\providecommand \@@endlink[0]{}%
\providecommand \url  [0]{\begingroup\@sanitize@url \@url }%
\providecommand \@url [1]{\endgroup\@href {#1}{\urlprefix }}%
\providecommand \urlprefix  [0]{URL }%
\providecommand \Eprint [0]{\href }%
\providecommand \doibase [0]{https://doi.org/}%
\providecommand \selectlanguage [0]{\@gobble}%
\providecommand \bibinfo  [0]{\@secondoftwo}%
\providecommand \bibfield  [0]{\@secondoftwo}%
\providecommand \translation [1]{[#1]}%
\providecommand \BibitemOpen [0]{}%
\providecommand \bibitemStop [0]{}%
\providecommand \bibitemNoStop [0]{.\EOS\space}%
\providecommand \EOS [0]{\spacefactor3000\relax}%
\providecommand \BibitemShut  [1]{\csname bibitem#1\endcsname}%
\let\auto@bib@innerbib\@empty
\bibitem [{\citenamefont {Lu}\ \emph {et~al.}(2014)\citenamefont {Lu},
  \citenamefont {Joannopoulos},\ and\ \citenamefont
  {Solja{\v{c}}i{\'c}}}]{lu2014}%
  \BibitemOpen
  \bibfield  {author} {\bibinfo {author} {\bibfnamefont {L.}~\bibnamefont
  {Lu}}, \bibinfo {author} {\bibfnamefont {J.~D.}\ \bibnamefont
  {Joannopoulos}},\ and\ \bibinfo {author} {\bibfnamefont {M.}~\bibnamefont
  {Solja{\v{c}}i{\'c}}},\ }\bibfield  {title} {\bibinfo {title} {Topological
  photonics},\ }\href@noop {} {\bibfield  {journal} {\bibinfo  {journal} {Nat.
  Photon.}\ }\textbf {\bibinfo {volume} {8}},\ \bibinfo {pages} {821} (\bibinfo
  {year} {2014})}\BibitemShut {NoStop}%
\bibitem [{\citenamefont {Khanikaev}\ and\ \citenamefont
  {Shvets}(2017)}]{khanikaev2017}%
  \BibitemOpen
  \bibfield  {author} {\bibinfo {author} {\bibfnamefont {A.~B.}\ \bibnamefont
  {Khanikaev}}\ and\ \bibinfo {author} {\bibfnamefont {G.}~\bibnamefont
  {Shvets}},\ }\bibfield  {title} {\bibinfo {title} {Two-dimensional
  topological photonics},\ }\href@noop {} {\bibfield  {journal} {\bibinfo
  {journal} {Nature Photon.}\ }\textbf {\bibinfo {volume} {11}},\ \bibinfo
  {pages} {763} (\bibinfo {year} {2017})}\BibitemShut {NoStop}%
\bibitem [{\citenamefont {Ozawa}\ \emph {et~al.}(2019)\citenamefont {Ozawa},
  \citenamefont {Price}, \citenamefont {Amo}, \citenamefont {Goldman},
  \citenamefont {Hafezi}, \citenamefont {Lu}, \citenamefont {Rechtsman},
  \citenamefont {Schuster}, \citenamefont {Simon}, \citenamefont {Zilberberg}
  \emph {et~al.}}]{ozawa2019}%
  \BibitemOpen
  \bibfield  {author} {\bibinfo {author} {\bibfnamefont {T.}~\bibnamefont
  {Ozawa}}, \bibinfo {author} {\bibfnamefont {H.~M.}\ \bibnamefont {Price}},
  \bibinfo {author} {\bibfnamefont {A.}~\bibnamefont {Amo}}, \bibinfo {author}
  {\bibfnamefont {N.}~\bibnamefont {Goldman}}, \bibinfo {author} {\bibfnamefont
  {M.}~\bibnamefont {Hafezi}}, \bibinfo {author} {\bibfnamefont
  {L.}~\bibnamefont {Lu}}, \bibinfo {author} {\bibfnamefont {M.~C.}\
  \bibnamefont {Rechtsman}}, \bibinfo {author} {\bibfnamefont {D.}~\bibnamefont
  {Schuster}}, \bibinfo {author} {\bibfnamefont {J.}~\bibnamefont {Simon}},
  \bibinfo {author} {\bibfnamefont {O.}~\bibnamefont {Zilberberg}}, \emph
  {et~al.},\ }\bibfield  {title} {\bibinfo {title} {Topological photonics},\
  }\href@noop {} {\bibfield  {journal} {\bibinfo  {journal} {Rev. Mod. Phys.}\
  }\textbf {\bibinfo {volume} {91}},\ \bibinfo {pages} {015006} (\bibinfo
  {year} {2019})}\BibitemShut {NoStop}%
\bibitem [{\citenamefont {Hasan}\ and\ \citenamefont {Kane}(2010)}]{hasan2010}%
  \BibitemOpen
  \bibfield  {author} {\bibinfo {author} {\bibfnamefont {M.~Z.}\ \bibnamefont
  {Hasan}}\ and\ \bibinfo {author} {\bibfnamefont {C.~L.}\ \bibnamefont
  {Kane}},\ }\bibfield  {title} {\bibinfo {title} {Colloquium: topological
  insulators},\ }\href@noop {} {\bibfield  {journal} {\bibinfo  {journal} {Rev.
  Mod. Phys.}\ }\textbf {\bibinfo {volume} {82}},\ \bibinfo {pages} {3045}
  (\bibinfo {year} {2010})}\BibitemShut {NoStop}%
\bibitem [{\citenamefont {Qi}\ and\ \citenamefont {Zhang}(2011)}]{qi2011}%
  \BibitemOpen
  \bibfield  {author} {\bibinfo {author} {\bibfnamefont {X.-L.}\ \bibnamefont
  {Qi}}\ and\ \bibinfo {author} {\bibfnamefont {S.-C.}\ \bibnamefont {Zhang}},\
  }\bibfield  {title} {\bibinfo {title} {Topological insulators and
  superconductors},\ }\href@noop {} {\bibfield  {journal} {\bibinfo  {journal}
  {Rev. Mod. Phys.}\ }\textbf {\bibinfo {volume} {83}},\ \bibinfo {pages}
  {1057} (\bibinfo {year} {2011})}\BibitemShut {NoStop}%
\bibitem [{\citenamefont {Haldane}\ and\ \citenamefont
  {Raghu}(2008)}]{haldane2008}%
  \BibitemOpen
  \bibfield  {author} {\bibinfo {author} {\bibfnamefont {F.}~\bibnamefont
  {Haldane}}\ and\ \bibinfo {author} {\bibfnamefont {S.}~\bibnamefont
  {Raghu}},\ }\bibfield  {title} {\bibinfo {title} {Possible realization of
  directional optical waveguides in photonic crystals with broken time-reversal
  symmetry},\ }\href@noop {} {\bibfield  {journal} {\bibinfo  {journal} {Phys.
  Rev. Lett.}\ }\textbf {\bibinfo {volume} {100}},\ \bibinfo {pages} {013904}
  (\bibinfo {year} {2008})}\BibitemShut {NoStop}%
\bibitem [{\citenamefont {Raghu}\ and\ \citenamefont
  {Haldane}(2008)}]{raghu2008}%
  \BibitemOpen
  \bibfield  {author} {\bibinfo {author} {\bibfnamefont {S.}~\bibnamefont
  {Raghu}}\ and\ \bibinfo {author} {\bibfnamefont {F.~D.~M.}\ \bibnamefont
  {Haldane}},\ }\bibfield  {title} {\bibinfo {title} {Analogs of
  quantum-\text{H}all-effect edge states in photonic crystals},\ }\href@noop {}
  {\bibfield  {journal} {\bibinfo  {journal} {Phys. Rev. A}\ }\textbf {\bibinfo
  {volume} {78}},\ \bibinfo {pages} {033834} (\bibinfo {year}
  {2008})}\BibitemShut {NoStop}%
\bibitem [{\citenamefont {Wang}\ \emph {et~al.}(2008)\citenamefont {Wang},
  \citenamefont {Chong}, \citenamefont {Joannopoulos},\ and\ \citenamefont
  {Solja{\v{c}}i{\'c}}}]{wang2008}%
  \BibitemOpen
  \bibfield  {author} {\bibinfo {author} {\bibfnamefont {Z.}~\bibnamefont
  {Wang}}, \bibinfo {author} {\bibfnamefont {Y.}~\bibnamefont {Chong}},
  \bibinfo {author} {\bibfnamefont {J.~D.}\ \bibnamefont {Joannopoulos}},\ and\
  \bibinfo {author} {\bibfnamefont {M.}~\bibnamefont {Solja{\v{c}}i{\'c}}},\
  }\bibfield  {title} {\bibinfo {title} {Reflection-free one-way edge modes in
  a gyromagnetic photonic crystal},\ }\href@noop {} {\bibfield  {journal}
  {\bibinfo  {journal} {Phys. Rev. Lett.}\ }\textbf {\bibinfo {volume} {100}},\
  \bibinfo {pages} {013905} (\bibinfo {year} {2008})}\BibitemShut {NoStop}%
\bibitem [{\citenamefont {Wang}\ \emph {et~al.}(2009)\citenamefont {Wang},
  \citenamefont {Chong}, \citenamefont {Joannopoulos},\ and\ \citenamefont
  {Solja{\v{c}}i{\'c}}}]{wang2009}%
  \BibitemOpen
  \bibfield  {author} {\bibinfo {author} {\bibfnamefont {Z.}~\bibnamefont
  {Wang}}, \bibinfo {author} {\bibfnamefont {Y.}~\bibnamefont {Chong}},
  \bibinfo {author} {\bibfnamefont {J.~D.}\ \bibnamefont {Joannopoulos}},\ and\
  \bibinfo {author} {\bibfnamefont {M.}~\bibnamefont {Solja{\v{c}}i{\'c}}},\
  }\bibfield  {title} {\bibinfo {title} {Observation of unidirectional
  backscattering-immune topological electromagnetic states},\ }\href@noop {}
  {\bibfield  {journal} {\bibinfo  {journal} {Nature}\ }\textbf {\bibinfo
  {volume} {461}},\ \bibinfo {pages} {772} (\bibinfo {year}
  {2009})}\BibitemShut {NoStop}%
\bibitem [{\citenamefont {Fang}\ \emph {et~al.}(2012)\citenamefont {Fang},
  \citenamefont {Yu},\ and\ \citenamefont {Fan}}]{fang2012}%
  \BibitemOpen
  \bibfield  {author} {\bibinfo {author} {\bibfnamefont {K.}~\bibnamefont
  {Fang}}, \bibinfo {author} {\bibfnamefont {Z.}~\bibnamefont {Yu}},\ and\
  \bibinfo {author} {\bibfnamefont {S.}~\bibnamefont {Fan}},\ }\bibfield
  {title} {\bibinfo {title} {Realizing effective magnetic field for photons by
  controlling the phase of dynamic modulation},\ }\href@noop {} {\bibfield
  {journal} {\bibinfo  {journal} {Nat. Photon.}\ }\textbf {\bibinfo {volume}
  {6}},\ \bibinfo {pages} {782} (\bibinfo {year} {2012})}\BibitemShut {NoStop}%
\bibitem [{\citenamefont {Rechtsman}\ \emph {et~al.}(2013)\citenamefont
  {Rechtsman}, \citenamefont {Zeuner}, \citenamefont {Plotnik}, \citenamefont
  {Lumer}, \citenamefont {Podolsky}, \citenamefont {Dreisow}, \citenamefont
  {Nolte}, \citenamefont {Segev},\ and\ \citenamefont
  {Szameit}}]{rechtsman2013}%
  \BibitemOpen
  \bibfield  {author} {\bibinfo {author} {\bibfnamefont {M.~C.}\ \bibnamefont
  {Rechtsman}}, \bibinfo {author} {\bibfnamefont {J.~M.}\ \bibnamefont
  {Zeuner}}, \bibinfo {author} {\bibfnamefont {Y.}~\bibnamefont {Plotnik}},
  \bibinfo {author} {\bibfnamefont {Y.}~\bibnamefont {Lumer}}, \bibinfo
  {author} {\bibfnamefont {D.}~\bibnamefont {Podolsky}}, \bibinfo {author}
  {\bibfnamefont {F.}~\bibnamefont {Dreisow}}, \bibinfo {author} {\bibfnamefont
  {S.}~\bibnamefont {Nolte}}, \bibinfo {author} {\bibfnamefont
  {M.}~\bibnamefont {Segev}},\ and\ \bibinfo {author} {\bibfnamefont
  {A.}~\bibnamefont {Szameit}},\ }\bibfield  {title} {\bibinfo {title}
  {Photonic \text{F}loquet topological insulators},\ }\href@noop {} {\bibfield
  {journal} {\bibinfo  {journal} {Nature}\ }\textbf {\bibinfo {volume} {496}},\
  \bibinfo {pages} {196} (\bibinfo {year} {2013})}\BibitemShut {NoStop}%
\bibitem [{\citenamefont {Liang}\ and\ \citenamefont
  {Chong}(2013)}]{liang2013}%
  \BibitemOpen
  \bibfield  {author} {\bibinfo {author} {\bibfnamefont {G.}~\bibnamefont
  {Liang}}\ and\ \bibinfo {author} {\bibfnamefont {Y.}~\bibnamefont {Chong}},\
  }\bibfield  {title} {\bibinfo {title} {Optical resonator analog of a
  two-dimensional topological insulator},\ }\href@noop {} {\bibfield  {journal}
  {\bibinfo  {journal} {Phys. Rev. Lett.}\ }\textbf {\bibinfo {volume} {110}},\
  \bibinfo {pages} {203904} (\bibinfo {year} {2013})}\BibitemShut {NoStop}%
\bibitem [{\citenamefont {Gao}\ \emph {et~al.}(2016)\citenamefont {Gao},
  \citenamefont {Gao}, \citenamefont {Shi}, \citenamefont {Yang}, \citenamefont
  {Lin}, \citenamefont {Xu}, \citenamefont {Joannopoulos}, \citenamefont
  {Solja{\v{c}}i{\'c}}, \citenamefont {Chen}, \citenamefont {Lu} \emph
  {et~al.}}]{gao2016}%
  \BibitemOpen
  \bibfield  {author} {\bibinfo {author} {\bibfnamefont {F.}~\bibnamefont
  {Gao}}, \bibinfo {author} {\bibfnamefont {Z.}~\bibnamefont {Gao}}, \bibinfo
  {author} {\bibfnamefont {X.}~\bibnamefont {Shi}}, \bibinfo {author}
  {\bibfnamefont {Z.}~\bibnamefont {Yang}}, \bibinfo {author} {\bibfnamefont
  {X.}~\bibnamefont {Lin}}, \bibinfo {author} {\bibfnamefont {H.}~\bibnamefont
  {Xu}}, \bibinfo {author} {\bibfnamefont {J.~D.}\ \bibnamefont
  {Joannopoulos}}, \bibinfo {author} {\bibfnamefont {M.}~\bibnamefont
  {Solja{\v{c}}i{\'c}}}, \bibinfo {author} {\bibfnamefont {H.}~\bibnamefont
  {Chen}}, \bibinfo {author} {\bibfnamefont {L.}~\bibnamefont {Lu}}, \emph
  {et~al.},\ }\bibfield  {title} {\bibinfo {title} {Probing topological
  protection using a designer surface plasmon structure},\ }\href@noop {}
  {\bibfield  {journal} {\bibinfo  {journal} {Nat. Commun.}\ }\textbf {\bibinfo
  {volume} {7}},\ \bibinfo {pages} {11619} (\bibinfo {year}
  {2016})}\BibitemShut {NoStop}%
\bibitem [{\citenamefont {Maczewsky}\ \emph {et~al.}(2017)\citenamefont
  {Maczewsky}, \citenamefont {Zeuner}, \citenamefont {Nolte},\ and\
  \citenamefont {Szameit}}]{maczewsky2017}%
  \BibitemOpen
  \bibfield  {author} {\bibinfo {author} {\bibfnamefont {L.~J.}\ \bibnamefont
  {Maczewsky}}, \bibinfo {author} {\bibfnamefont {J.~M.}\ \bibnamefont
  {Zeuner}}, \bibinfo {author} {\bibfnamefont {S.}~\bibnamefont {Nolte}},\ and\
  \bibinfo {author} {\bibfnamefont {A.}~\bibnamefont {Szameit}},\ }\bibfield
  {title} {\bibinfo {title} {Observation of photonic anomalous \text{F}loquet
  topological insulators},\ }\href@noop {} {\bibfield  {journal} {\bibinfo
  {journal} {Nat. Commun.}\ }\textbf {\bibinfo {volume} {8}},\ \bibinfo {pages}
  {13756} (\bibinfo {year} {2017})}\BibitemShut {NoStop}%
\bibitem [{\citenamefont {Mukherjee}\ \emph {et~al.}(2017)\citenamefont
  {Mukherjee}, \citenamefont {Spracklen}, \citenamefont {Valiente},
  \citenamefont {Andersson}, \citenamefont {{\"O}hberg}, \citenamefont
  {Goldman},\ and\ \citenamefont {Thomson}}]{mukherjee2017}%
  \BibitemOpen
  \bibfield  {author} {\bibinfo {author} {\bibfnamefont {S.}~\bibnamefont
  {Mukherjee}}, \bibinfo {author} {\bibfnamefont {A.}~\bibnamefont
  {Spracklen}}, \bibinfo {author} {\bibfnamefont {M.}~\bibnamefont {Valiente}},
  \bibinfo {author} {\bibfnamefont {E.}~\bibnamefont {Andersson}}, \bibinfo
  {author} {\bibfnamefont {P.}~\bibnamefont {{\"O}hberg}}, \bibinfo {author}
  {\bibfnamefont {N.}~\bibnamefont {Goldman}},\ and\ \bibinfo {author}
  {\bibfnamefont {R.~R.}\ \bibnamefont {Thomson}},\ }\bibfield  {title}
  {\bibinfo {title} {Experimental observation of anomalous topological edge
  modes in a slowly driven photonic lattice},\ }\href@noop {} {\bibfield
  {journal} {\bibinfo  {journal} {Nat. Commun.}\ }\textbf {\bibinfo {volume}
  {8}},\ \bibinfo {pages} {13918} (\bibinfo {year} {2017})}\BibitemShut
  {NoStop}%
\bibitem [{\citenamefont {Hafezi}\ \emph {et~al.}(2011)\citenamefont {Hafezi},
  \citenamefont {Demler}, \citenamefont {Lukin},\ and\ \citenamefont
  {Taylor}}]{hafezi2011}%
  \BibitemOpen
  \bibfield  {author} {\bibinfo {author} {\bibfnamefont {M.}~\bibnamefont
  {Hafezi}}, \bibinfo {author} {\bibfnamefont {E.~A.}\ \bibnamefont {Demler}},
  \bibinfo {author} {\bibfnamefont {M.~D.}\ \bibnamefont {Lukin}},\ and\
  \bibinfo {author} {\bibfnamefont {J.~M.}\ \bibnamefont {Taylor}},\ }\bibfield
   {title} {\bibinfo {title} {Robust optical delay lines with topological
  protection},\ }\href@noop {} {\bibfield  {journal} {\bibinfo  {journal} {Nat.
  Phys.}\ }\textbf {\bibinfo {volume} {7}},\ \bibinfo {pages} {907} (\bibinfo
  {year} {2011})}\BibitemShut {NoStop}%
\bibitem [{\citenamefont {Hafezi}\ \emph {et~al.}(2013)\citenamefont {Hafezi},
  \citenamefont {Mittal}, \citenamefont {Fan}, \citenamefont {Migdall},\ and\
  \citenamefont {Taylor}}]{hafezi2013}%
  \BibitemOpen
  \bibfield  {author} {\bibinfo {author} {\bibfnamefont {M.}~\bibnamefont
  {Hafezi}}, \bibinfo {author} {\bibfnamefont {S.}~\bibnamefont {Mittal}},
  \bibinfo {author} {\bibfnamefont {J.}~\bibnamefont {Fan}}, \bibinfo {author}
  {\bibfnamefont {A.}~\bibnamefont {Migdall}},\ and\ \bibinfo {author}
  {\bibfnamefont {J.}~\bibnamefont {Taylor}},\ }\bibfield  {title} {\bibinfo
  {title} {Imaging topological edge states in silicon photonics},\ }\href@noop
  {} {\bibfield  {journal} {\bibinfo  {journal} {Nat. Photon.}\ }\textbf
  {\bibinfo {volume} {7}},\ \bibinfo {pages} {1001} (\bibinfo {year}
  {2013})}\BibitemShut {NoStop}%
\bibitem [{\citenamefont {Khanikaev}\ \emph {et~al.}(2013)\citenamefont
  {Khanikaev}, \citenamefont {Mousavi}, \citenamefont {Tse}, \citenamefont
  {Kargarian}, \citenamefont {MacDonald},\ and\ \citenamefont
  {Shvets}}]{khanikaev2013}%
  \BibitemOpen
  \bibfield  {author} {\bibinfo {author} {\bibfnamefont {A.~B.}\ \bibnamefont
  {Khanikaev}}, \bibinfo {author} {\bibfnamefont {S.~H.}\ \bibnamefont
  {Mousavi}}, \bibinfo {author} {\bibfnamefont {W.-K.}\ \bibnamefont {Tse}},
  \bibinfo {author} {\bibfnamefont {M.}~\bibnamefont {Kargarian}}, \bibinfo
  {author} {\bibfnamefont {A.~H.}\ \bibnamefont {MacDonald}},\ and\ \bibinfo
  {author} {\bibfnamefont {G.}~\bibnamefont {Shvets}},\ }\bibfield  {title}
  {\bibinfo {title} {Photonic topological insulators},\ }\href@noop {}
  {\bibfield  {journal} {\bibinfo  {journal} {Nat. Mater.}\ }\textbf {\bibinfo
  {volume} {12}},\ \bibinfo {pages} {233} (\bibinfo {year} {2013})}\BibitemShut
  {NoStop}%
\bibitem [{\citenamefont {Chen}\ \emph {et~al.}(2014)\citenamefont {Chen},
  \citenamefont {Jiang}, \citenamefont {Chen}, \citenamefont {Zhu},
  \citenamefont {Zhou}, \citenamefont {Dong},\ and\ \citenamefont
  {Chan}}]{chen2014}%
  \BibitemOpen
  \bibfield  {author} {\bibinfo {author} {\bibfnamefont {W.-J.}\ \bibnamefont
  {Chen}}, \bibinfo {author} {\bibfnamefont {S.-J.}\ \bibnamefont {Jiang}},
  \bibinfo {author} {\bibfnamefont {X.-D.}\ \bibnamefont {Chen}}, \bibinfo
  {author} {\bibfnamefont {B.}~\bibnamefont {Zhu}}, \bibinfo {author}
  {\bibfnamefont {L.}~\bibnamefont {Zhou}}, \bibinfo {author} {\bibfnamefont
  {J.-W.}\ \bibnamefont {Dong}},\ and\ \bibinfo {author} {\bibfnamefont
  {C.~T.}\ \bibnamefont {Chan}},\ }\bibfield  {title} {\bibinfo {title}
  {Experimental realization of photonic topological insulator in a uniaxial
  metacrystal waveguide},\ }\href@noop {} {\bibfield  {journal} {\bibinfo
  {journal} {Nat. Commun.}\ }\textbf {\bibinfo {volume} {5}},\ \bibinfo {pages}
  {5782} (\bibinfo {year} {2014})}\BibitemShut {NoStop}%
\bibitem [{\citenamefont {Wu}\ and\ \citenamefont {Hu}(2015)}]{wu2015}%
  \BibitemOpen
  \bibfield  {author} {\bibinfo {author} {\bibfnamefont {L.-H.}\ \bibnamefont
  {Wu}}\ and\ \bibinfo {author} {\bibfnamefont {X.}~\bibnamefont {Hu}},\
  }\bibfield  {title} {\bibinfo {title} {Scheme for achieving a topological
  photonic crystal by using dielectric material},\ }\href@noop {} {\bibfield
  {journal} {\bibinfo  {journal} {Phys. Rev. Lett.}\ }\textbf {\bibinfo
  {volume} {114}},\ \bibinfo {pages} {223901} (\bibinfo {year}
  {2015})}\BibitemShut {NoStop}%
\bibitem [{\citenamefont {Cheng}\ \emph {et~al.}(2016)\citenamefont {Cheng},
  \citenamefont {Jouvaud}, \citenamefont {Ni}, \citenamefont {Mousavi},
  \citenamefont {Genack},\ and\ \citenamefont {Khanikaev}}]{cheng2016}%
  \BibitemOpen
  \bibfield  {author} {\bibinfo {author} {\bibfnamefont {X.}~\bibnamefont
  {Cheng}}, \bibinfo {author} {\bibfnamefont {C.}~\bibnamefont {Jouvaud}},
  \bibinfo {author} {\bibfnamefont {X.}~\bibnamefont {Ni}}, \bibinfo {author}
  {\bibfnamefont {S.~H.}\ \bibnamefont {Mousavi}}, \bibinfo {author}
  {\bibfnamefont {A.~Z.}\ \bibnamefont {Genack}},\ and\ \bibinfo {author}
  {\bibfnamefont {A.~B.}\ \bibnamefont {Khanikaev}},\ }\bibfield  {title}
  {\bibinfo {title} {Robust reconfigurable electromagnetic pathways within a
  photonic topological insulator},\ }\href@noop {} {\bibfield  {journal}
  {\bibinfo  {journal} {Nat. Mater.}\ }\textbf {\bibinfo {volume} {15}},\
  \bibinfo {pages} {542} (\bibinfo {year} {2016})}\BibitemShut {NoStop}%
\bibitem [{\citenamefont {Ma}\ and\ \citenamefont {Shvets}(2016)}]{ma2016}%
  \BibitemOpen
  \bibfield  {author} {\bibinfo {author} {\bibfnamefont {T.}~\bibnamefont
  {Ma}}\ and\ \bibinfo {author} {\bibfnamefont {G.}~\bibnamefont {Shvets}},\
  }\bibfield  {title} {\bibinfo {title} {All-\text{S}i valley-\text{H}all
  photonic topological insulator},\ }\href@noop {} {\bibfield  {journal}
  {\bibinfo  {journal} {New J. Phys.}\ }\textbf {\bibinfo {volume} {18}},\
  \bibinfo {pages} {025012} (\bibinfo {year} {2016})}\BibitemShut {NoStop}%
\bibitem [{\citenamefont {Dong}\ \emph {et~al.}(2017)\citenamefont {Dong},
  \citenamefont {Chen}, \citenamefont {Zhu}, \citenamefont {Wang},\ and\
  \citenamefont {Zhang}}]{dong2017}%
  \BibitemOpen
  \bibfield  {author} {\bibinfo {author} {\bibfnamefont {J.-W.}\ \bibnamefont
  {Dong}}, \bibinfo {author} {\bibfnamefont {X.-D.}\ \bibnamefont {Chen}},
  \bibinfo {author} {\bibfnamefont {H.}~\bibnamefont {Zhu}}, \bibinfo {author}
  {\bibfnamefont {Y.}~\bibnamefont {Wang}},\ and\ \bibinfo {author}
  {\bibfnamefont {X.}~\bibnamefont {Zhang}},\ }\bibfield  {title} {\bibinfo
  {title} {Valley photonic crystals for control of spin and topology},\
  }\href@noop {} {\bibfield  {journal} {\bibinfo  {journal} {Nat. Mater.}\
  }\textbf {\bibinfo {volume} {16}},\ \bibinfo {pages} {298} (\bibinfo {year}
  {2017})}\BibitemShut {NoStop}%
\bibitem [{\citenamefont {Wu}\ \emph {et~al.}(2017)\citenamefont {Wu},
  \citenamefont {Meng}, \citenamefont {Tian}, \citenamefont {Huang},
  \citenamefont {Xiang}, \citenamefont {Han},\ and\ \citenamefont
  {Wen}}]{wu2017}%
  \BibitemOpen
  \bibfield  {author} {\bibinfo {author} {\bibfnamefont {X.}~\bibnamefont
  {Wu}}, \bibinfo {author} {\bibfnamefont {Y.}~\bibnamefont {Meng}}, \bibinfo
  {author} {\bibfnamefont {J.}~\bibnamefont {Tian}}, \bibinfo {author}
  {\bibfnamefont {Y.}~\bibnamefont {Huang}}, \bibinfo {author} {\bibfnamefont
  {H.}~\bibnamefont {Xiang}}, \bibinfo {author} {\bibfnamefont
  {D.}~\bibnamefont {Han}},\ and\ \bibinfo {author} {\bibfnamefont
  {W.}~\bibnamefont {Wen}},\ }\bibfield  {title} {\bibinfo {title} {Direct
  observation of valley-polarized topological edge states in designer surface
  plasmon crystals},\ }\href@noop {} {\bibfield  {journal} {\bibinfo  {journal}
  {Nat. Commun.}\ }\textbf {\bibinfo {volume} {8}},\ \bibinfo {pages} {1304}
  (\bibinfo {year} {2017})}\BibitemShut {NoStop}%
\bibitem [{\citenamefont {Gao}\ \emph {et~al.}(2018)\citenamefont {Gao},
  \citenamefont {Xue}, \citenamefont {Yang}, \citenamefont {Lai}, \citenamefont
  {Yu}, \citenamefont {Lin}, \citenamefont {Chong}, \citenamefont {Shvets},\
  and\ \citenamefont {Zhang}}]{gao2018}%
  \BibitemOpen
  \bibfield  {author} {\bibinfo {author} {\bibfnamefont {F.}~\bibnamefont
  {Gao}}, \bibinfo {author} {\bibfnamefont {H.}~\bibnamefont {Xue}}, \bibinfo
  {author} {\bibfnamefont {Z.}~\bibnamefont {Yang}}, \bibinfo {author}
  {\bibfnamefont {K.}~\bibnamefont {Lai}}, \bibinfo {author} {\bibfnamefont
  {Y.}~\bibnamefont {Yu}}, \bibinfo {author} {\bibfnamefont {X.}~\bibnamefont
  {Lin}}, \bibinfo {author} {\bibfnamefont {Y.}~\bibnamefont {Chong}}, \bibinfo
  {author} {\bibfnamefont {G.}~\bibnamefont {Shvets}},\ and\ \bibinfo {author}
  {\bibfnamefont {B.}~\bibnamefont {Zhang}},\ }\bibfield  {title} {\bibinfo
  {title} {Topologically protected refraction of robust kink states in valley
  photonic crystals},\ }\href@noop {} {\bibfield  {journal} {\bibinfo
  {journal} {Nat. Phys.}\ }\textbf {\bibinfo {volume} {14}},\ \bibinfo {pages}
  {140} (\bibinfo {year} {2018})}\BibitemShut {NoStop}%
\bibitem [{\citenamefont {Noh}\ \emph {et~al.}(2018)\citenamefont {Noh},
  \citenamefont {Huang}, \citenamefont {Chen},\ and\ \citenamefont
  {Rechtsman}}]{noh2018}%
  \BibitemOpen
  \bibfield  {author} {\bibinfo {author} {\bibfnamefont {J.}~\bibnamefont
  {Noh}}, \bibinfo {author} {\bibfnamefont {S.}~\bibnamefont {Huang}}, \bibinfo
  {author} {\bibfnamefont {K.~P.}\ \bibnamefont {Chen}},\ and\ \bibinfo
  {author} {\bibfnamefont {M.~C.}\ \bibnamefont {Rechtsman}},\ }\bibfield
  {title} {\bibinfo {title} {Observation of photonic topological valley hall
  edge states},\ }\href@noop {} {\bibfield  {journal} {\bibinfo  {journal}
  {Phys. Rev. Lett.}\ }\textbf {\bibinfo {volume} {120}},\ \bibinfo {pages}
  {063902} (\bibinfo {year} {2018})}\BibitemShut {NoStop}%
\bibitem [{\citenamefont {Lu}\ \emph {et~al.}(2013)\citenamefont {Lu},
  \citenamefont {Fu}, \citenamefont {Joannopoulos},\ and\ \citenamefont
  {Solja{\v{c}}i{\'c}}}]{lu2013}%
  \BibitemOpen
  \bibfield  {author} {\bibinfo {author} {\bibfnamefont {L.}~\bibnamefont
  {Lu}}, \bibinfo {author} {\bibfnamefont {L.}~\bibnamefont {Fu}}, \bibinfo
  {author} {\bibfnamefont {J.~D.}\ \bibnamefont {Joannopoulos}},\ and\ \bibinfo
  {author} {\bibfnamefont {M.}~\bibnamefont {Solja{\v{c}}i{\'c}}},\ }\bibfield
  {title} {\bibinfo {title} {Weyl points and line nodes in gyroid photonic
  crystals},\ }\href@noop {} {\bibfield  {journal} {\bibinfo  {journal} {Nat.
  Photon.}\ }\textbf {\bibinfo {volume} {7}},\ \bibinfo {pages} {294} (\bibinfo
  {year} {2013})}\BibitemShut {NoStop}%
\bibitem [{\citenamefont {Lu}\ \emph {et~al.}(2015)\citenamefont {Lu},
  \citenamefont {Wang}, \citenamefont {Ye}, \citenamefont {Ran}, \citenamefont
  {Fu}, \citenamefont {Joannopoulos},\ and\ \citenamefont
  {Solja{\v{c}}i{\'c}}}]{lu2015}%
  \BibitemOpen
  \bibfield  {author} {\bibinfo {author} {\bibfnamefont {L.}~\bibnamefont
  {Lu}}, \bibinfo {author} {\bibfnamefont {Z.}~\bibnamefont {Wang}}, \bibinfo
  {author} {\bibfnamefont {D.}~\bibnamefont {Ye}}, \bibinfo {author}
  {\bibfnamefont {L.}~\bibnamefont {Ran}}, \bibinfo {author} {\bibfnamefont
  {L.}~\bibnamefont {Fu}}, \bibinfo {author} {\bibfnamefont {J.~D.}\
  \bibnamefont {Joannopoulos}},\ and\ \bibinfo {author} {\bibfnamefont
  {M.}~\bibnamefont {Solja{\v{c}}i{\'c}}},\ }\bibfield  {title} {\bibinfo
  {title} {Experimental observation of \text{W}eyl points},\ }\href@noop {}
  {\bibfield  {journal} {\bibinfo  {journal} {Science}\ }\textbf {\bibinfo
  {volume} {349}},\ \bibinfo {pages} {622} (\bibinfo {year}
  {2015})}\BibitemShut {NoStop}%
\bibitem [{\citenamefont {Baba}(2008)}]{baba2008}%
  \BibitemOpen
  \bibfield  {author} {\bibinfo {author} {\bibfnamefont {T.}~\bibnamefont
  {Baba}},\ }\bibfield  {title} {\bibinfo {title} {Slow light in photonic
  crystals},\ }\href@noop {} {\bibfield  {journal} {\bibinfo  {journal} {Nat.
  Photon.}\ }\textbf {\bibinfo {volume} {2}},\ \bibinfo {pages} {465} (\bibinfo
  {year} {2008})}\BibitemShut {NoStop}%
\bibitem [{\citenamefont {Melati}\ \emph {et~al.}(2014)\citenamefont {Melati},
  \citenamefont {Melloni},\ and\ \citenamefont {Morichetti}}]{melati2014}%
  \BibitemOpen
  \bibfield  {author} {\bibinfo {author} {\bibfnamefont {D.}~\bibnamefont
  {Melati}}, \bibinfo {author} {\bibfnamefont {A.}~\bibnamefont {Melloni}},\
  and\ \bibinfo {author} {\bibfnamefont {F.}~\bibnamefont {Morichetti}},\
  }\bibfield  {title} {\bibinfo {title} {Real photonic waveguides: guiding
  light through imperfections},\ }\href@noop {} {\bibfield  {journal} {\bibinfo
   {journal} {Adv. Opt. Photon.}\ }\textbf {\bibinfo {volume} {6}},\ \bibinfo
  {pages} {156} (\bibinfo {year} {2014})}\BibitemShut {NoStop}%
\bibitem [{\citenamefont {Thouless}\ \emph {et~al.}(1982)\citenamefont
  {Thouless}, \citenamefont {Kohmoto}, \citenamefont {Nightingale},\ and\
  \citenamefont {den Nijs}}]{thouless1982}%
  \BibitemOpen
  \bibfield  {author} {\bibinfo {author} {\bibfnamefont {D.~J.}\ \bibnamefont
  {Thouless}}, \bibinfo {author} {\bibfnamefont {M.}~\bibnamefont {Kohmoto}},
  \bibinfo {author} {\bibfnamefont {M.~P.}\ \bibnamefont {Nightingale}},\ and\
  \bibinfo {author} {\bibfnamefont {M.}~\bibnamefont {den Nijs}},\ }\bibfield
  {title} {\bibinfo {title} {Quantized \text{H}all conductance in a
  two-dimensional periodic potential},\ }\href@noop {} {\bibfield  {journal}
  {\bibinfo  {journal} {Phys. Rev. Lett.}\ }\textbf {\bibinfo {volume} {49}},\
  \bibinfo {pages} {405} (\bibinfo {year} {1982})}\BibitemShut {NoStop}%
\bibitem [{\citenamefont {Yang}\ \emph {et~al.}(2013)\citenamefont {Yang},
  \citenamefont {Poo}, \citenamefont {Wu}, \citenamefont {Gu},\ and\
  \citenamefont {Chen}}]{yang2013}%
  \BibitemOpen
  \bibfield  {author} {\bibinfo {author} {\bibfnamefont {Y.}~\bibnamefont
  {Yang}}, \bibinfo {author} {\bibfnamefont {Y.}~\bibnamefont {Poo}}, \bibinfo
  {author} {\bibfnamefont {R.-x.}\ \bibnamefont {Wu}}, \bibinfo {author}
  {\bibfnamefont {Y.}~\bibnamefont {Gu}},\ and\ \bibinfo {author}
  {\bibfnamefont {P.}~\bibnamefont {Chen}},\ }\bibfield  {title} {\bibinfo
  {title} {Experimental demonstration of one-way slow wave in waveguide
  involving gyromagnetic photonic crystals},\ }\href@noop {} {\bibfield
  {journal} {\bibinfo  {journal} {Appl. Phys. Lett.}\ }\textbf {\bibinfo
  {volume} {102}},\ \bibinfo {pages} {231113} (\bibinfo {year}
  {2013})}\BibitemShut {NoStop}%
\bibitem [{\citenamefont {Lan}\ \emph {et~al.}(2020)\citenamefont {Lan},
  \citenamefont {You},\ and\ \citenamefont {Panoiu}}]{lan2020}%
  \BibitemOpen
  \bibfield  {author} {\bibinfo {author} {\bibfnamefont {Z.}~\bibnamefont
  {Lan}}, \bibinfo {author} {\bibfnamefont {J.~W.}\ \bibnamefont {You}},\ and\
  \bibinfo {author} {\bibfnamefont {N.~C.}\ \bibnamefont {Panoiu}},\ }\bibfield
   {title} {\bibinfo {title} {Nonlinear one-way edge-mode interactions for
  frequency mixing in topological photonic crystals},\ }\href@noop {}
  {\bibfield  {journal} {\bibinfo  {journal} {Phys. Rev. B}\ }\textbf {\bibinfo
  {volume} {101}},\ \bibinfo {pages} {155422} (\bibinfo {year}
  {2020})}\BibitemShut {NoStop}%
\bibitem [{\citenamefont {Guglielmon}\ and\ \citenamefont
  {Rechtsman}(2019)}]{guglielmon2019}%
  \BibitemOpen
  \bibfield  {author} {\bibinfo {author} {\bibfnamefont {J.}~\bibnamefont
  {Guglielmon}}\ and\ \bibinfo {author} {\bibfnamefont {M.~C.}\ \bibnamefont
  {Rechtsman}},\ }\bibfield  {title} {\bibinfo {title} {Broadband topological
  slow light through higher momentum-space winding},\ }\href@noop {} {\bibfield
   {journal} {\bibinfo  {journal} {Phys. Rev. Lett.}\ }\textbf {\bibinfo
  {volume} {122}},\ \bibinfo {pages} {153904} (\bibinfo {year}
  {2019})}\BibitemShut {NoStop}%
\bibitem [{\citenamefont {Poo}\ \emph {et~al.}(2011)\citenamefont {Poo},
  \citenamefont {Wu}, \citenamefont {Lin}, \citenamefont {Yang},\ and\
  \citenamefont {Chan}}]{poo2011}%
  \BibitemOpen
  \bibfield  {author} {\bibinfo {author} {\bibfnamefont {Y.}~\bibnamefont
  {Poo}}, \bibinfo {author} {\bibfnamefont {R.-x.}\ \bibnamefont {Wu}},
  \bibinfo {author} {\bibfnamefont {Z.}~\bibnamefont {Lin}}, \bibinfo {author}
  {\bibfnamefont {Y.}~\bibnamefont {Yang}},\ and\ \bibinfo {author}
  {\bibfnamefont {C.}~\bibnamefont {Chan}},\ }\bibfield  {title} {\bibinfo
  {title} {Experimental realization of self-guiding unidirectional
  electromagnetic edge states},\ }\href@noop {} {\bibfield  {journal} {\bibinfo
   {journal} {Phys. Rev. Lett.}\ }\textbf {\bibinfo {volume} {106}},\ \bibinfo
  {pages} {093903} (\bibinfo {year} {2011})}\BibitemShut {NoStop}%
\bibitem [{SM()}]{SM}%
  \BibitemOpen
  \href@noop {} {}\bibinfo {note} {See online Supplemental
  Material.}\BibitemShut {Stop}%
\bibitem [{\citenamefont {Yang}\ \emph {et~al.}(2015)\citenamefont {Yang},
  \citenamefont {Gao}, \citenamefont {Shi}, \citenamefont {Lin}, \citenamefont
  {Gao}, \citenamefont {Chong},\ and\ \citenamefont {Zhang}}]{yang2015}%
  \BibitemOpen
  \bibfield  {author} {\bibinfo {author} {\bibfnamefont {Z.}~\bibnamefont
  {Yang}}, \bibinfo {author} {\bibfnamefont {F.}~\bibnamefont {Gao}}, \bibinfo
  {author} {\bibfnamefont {X.}~\bibnamefont {Shi}}, \bibinfo {author}
  {\bibfnamefont {X.}~\bibnamefont {Lin}}, \bibinfo {author} {\bibfnamefont
  {Z.}~\bibnamefont {Gao}}, \bibinfo {author} {\bibfnamefont {Y.}~\bibnamefont
  {Chong}},\ and\ \bibinfo {author} {\bibfnamefont {B.}~\bibnamefont {Zhang}},\
  }\bibfield  {title} {\bibinfo {title} {Topological acoustics},\ }\href@noop
  {} {\bibfield  {journal} {\bibinfo  {journal} {Phys. Rev. Lett.}\ }\textbf
  {\bibinfo {volume} {114}},\ \bibinfo {pages} {114301} (\bibinfo {year}
  {2015})}\BibitemShut {NoStop}%
\bibitem [{\citenamefont {Ni}\ \emph {et~al.}(2015)\citenamefont {Ni},
  \citenamefont {He}, \citenamefont {Sun}, \citenamefont {Liu}, \citenamefont
  {Lu}, \citenamefont {Feng},\ and\ \citenamefont {Chen}}]{ni2015}%
  \BibitemOpen
  \bibfield  {author} {\bibinfo {author} {\bibfnamefont {X.}~\bibnamefont
  {Ni}}, \bibinfo {author} {\bibfnamefont {C.}~\bibnamefont {He}}, \bibinfo
  {author} {\bibfnamefont {X.-C.}\ \bibnamefont {Sun}}, \bibinfo {author}
  {\bibfnamefont {X.-p.}\ \bibnamefont {Liu}}, \bibinfo {author} {\bibfnamefont
  {M.-H.}\ \bibnamefont {Lu}}, \bibinfo {author} {\bibfnamefont
  {L.}~\bibnamefont {Feng}},\ and\ \bibinfo {author} {\bibfnamefont {Y.-F.}\
  \bibnamefont {Chen}},\ }\bibfield  {title} {\bibinfo {title} {Topologically
  protected one-way edge mode in networks of acoustic resonators with
  circulating air flow},\ }\href@noop {} {\bibfield  {journal} {\bibinfo
  {journal} {New J.Phys.}\ }\textbf {\bibinfo {volume} {17}},\ \bibinfo {pages}
  {053016} (\bibinfo {year} {2015})}\BibitemShut {NoStop}%
\bibitem [{\citenamefont {Khanikaev}\ \emph {et~al.}(2015)\citenamefont
  {Khanikaev}, \citenamefont {Fleury}, \citenamefont {Mousavi},\ and\
  \citenamefont {Alu}}]{khanikaev2015}%
  \BibitemOpen
  \bibfield  {author} {\bibinfo {author} {\bibfnamefont {A.~B.}\ \bibnamefont
  {Khanikaev}}, \bibinfo {author} {\bibfnamefont {R.}~\bibnamefont {Fleury}},
  \bibinfo {author} {\bibfnamefont {S.~H.}\ \bibnamefont {Mousavi}},\ and\
  \bibinfo {author} {\bibfnamefont {A.}~\bibnamefont {Alu}},\ }\bibfield
  {title} {\bibinfo {title} {Topologically robust sound propagation in an
  angular-momentum-biased graphene-like resonator lattice},\ }\href@noop {}
  {\bibfield  {journal} {\bibinfo  {journal} {Nat. Commun.}\ }\textbf {\bibinfo
  {volume} {6}},\ \bibinfo {pages} {8260} (\bibinfo {year} {2015})}\BibitemShut
  {NoStop}%
\bibitem [{\citenamefont {Ding}\ \emph {et~al.}(2019)\citenamefont {Ding},
  \citenamefont {Peng}, \citenamefont {Zhu}, \citenamefont {Fan}, \citenamefont
  {Yang}, \citenamefont {Liang}, \citenamefont {Zhu}, \citenamefont {Wan},\
  and\ \citenamefont {Cheng}}]{ding2019}%
  \BibitemOpen
  \bibfield  {author} {\bibinfo {author} {\bibfnamefont {Y.}~\bibnamefont
  {Ding}}, \bibinfo {author} {\bibfnamefont {Y.}~\bibnamefont {Peng}}, \bibinfo
  {author} {\bibfnamefont {Y.}~\bibnamefont {Zhu}}, \bibinfo {author}
  {\bibfnamefont {X.}~\bibnamefont {Fan}}, \bibinfo {author} {\bibfnamefont
  {J.}~\bibnamefont {Yang}}, \bibinfo {author} {\bibfnamefont {B.}~\bibnamefont
  {Liang}}, \bibinfo {author} {\bibfnamefont {X.}~\bibnamefont {Zhu}}, \bibinfo
  {author} {\bibfnamefont {X.}~\bibnamefont {Wan}},\ and\ \bibinfo {author}
  {\bibfnamefont {J.}~\bibnamefont {Cheng}},\ }\bibfield  {title} {\bibinfo
  {title} {Experimental demonstration of acoustic \text{C}hern insulators},\
  }\href@noop {} {\bibfield  {journal} {\bibinfo  {journal} {Phys. Rev. Lett.}\
  }\textbf {\bibinfo {volume} {122}},\ \bibinfo {pages} {014302} (\bibinfo
  {year} {2019})}\BibitemShut {NoStop}%
\end{thebibliography}
\end{document}